\documentclass[useAMS,usenatbib]{mn2e}

\input psfig.tex

\title[Extending the Butcher--Oemler effect up to
$z\sim0.7$]
{Extending the Butcher--Oemler effect up to
$z\sim0.7$\thanks{Based on observations obtained at the 3.6m ESO,
Canada--France--Hawaii, Cerro--Tololo and Kitt Peak telescopes.}}

\author[S. Andreon, C. Lobo, A. Iovino]
{S.~Andreon,$^1$\thanks{email: andreon@brera.mi.astro.it} 
C.~Lobo,$^{2,3}$ A.~Iovino,$^1$ \\ 
$^1$INAF--Osservatorio Astronomico di Brera, Milano, Italy\\
$^2$Departamento de Matem\'atica Aplicada, Faculdade de
    Ci\^encias da Universidade do Porto, Portugal\\
$^3$Centro de Astrof\'{\i}sica da Universidade do Porto, 
Rua das Estrelas, 4150-762 Porto, Portugal}
\date{Accepted ... Received ...}

\pagerange{\pageref{firstpage}--\pageref{lastpage}}
\pubyear{2003}

\begin{document}

\maketitle

\label{firstpage}

\begin{abstract} 
We have observed three clusters at $z\sim0.7$,
of richness comparable to the low redshift sample of Butcher \& Oemler
(BO), and determined their fraction of blue galaxies. 
When adopting the standard error definition, two clusters have a
low blue fraction for their redshifts, whereas the fraction of the
third one is compatible with the
expected value.  A detailed analysis of previous BO--like studies
that adopted different definitions of the
blue fraction shows that the modified definitions are affected by
contaminating signals:
colour segregation in clusters affects blue fractions
derived in fixed metric apertures, differential evolution of early
and late type spirals potentially affects blue fractions derived
with a non standard choice of the colour cut, the younger age of the
Universe at high redshift affects blue fractions computed with a
colour cut taken relatively to a fixed non evolving colour. Adopting
these definitions we find largely varying blue fractions. 
This thorough analysis
of the drawbacks of the different possible definitions of the blue
fraction should allow future studies to perform measures in the
same scale.
Finally, if one adopts a more refined error analysis to deal
with BO and our data,
a constant blue fraction with redshift cannot be excluded, showing that
the BO effect is still far from being detected beyond doubt.
\end{abstract}


\begin{table*}
\caption{Observations}
\begin{tabular}{lccccccccc}
\hline
Cluster name&  short name& 
 \multicolumn{4}{c}{exposure time [s]} & seeing & pixel size & observed & I
 completeness mag \\
\cline{3-6}
& & B & V & R & I & [arcsec] & [arcsec] & in run \\
\hline
J004831.6-294206.6 & cl0048-2942  & 2$\times$1920 & 2$\times$960 & 2$\times$600 & 2$\times$720 & 1.4 -- 1.6 & 0.314 & Aug '99 & 23.0\\
J224932.1-395804.6 & cl2245-3954  & 2$\times$1920 & 2$\times$960 & 2$\times$360 & 2$\times$480 & 0.7 -- 0.9 & 0.157 & Oct '98 & 22.8\\
J224513.2-395409.9 & cl2249-3958  & 2$\times$1920 & 2$\times$960 & 2$\times$360 & 2$\times$480 & 0.7 -- 0.9 & 0.157 & Oct '98 & 23.2\\
\hline
\end{tabular}
\label{tab:data}
\end{table*}

\section{Introduction}

Butcher \& Oemler (1978, 1984; BO hereafter) provided the first
dramatically clear evidence that galaxy populations differ at high and
low redshifts: clusters at high redshift contain a larger fraction of
blue galaxies than their nearby counterparts. Dressler et al. (1994)
have shown, by using images from the refurbished {\it Hubble Space
Telescope}, that the blue galaxies responsible for the
Butcher--Oemler effect in the
particular case of cluster cl 0939$+$4713 at $z = 0.41$ are late-type
spiral and irregular galaxies. Rakos \& Schombert (1995) found that
80\% of the galaxies in clusters at $z=0.9$ are blue, in
clear contrast to 20\% at $z = 0.4$. 

The abrupt variation in cluster colour content observed by Rakos \&
Schombert (1995) poses the problem of finding a highly efficient
mechanism that can account for these galaxy transformations on such
short time scales. In fact, the authors comment on the difficulty to
imagine a scenario where over 80\% of the cluster population is
destroyed or faded, especially since no remaining evidence (some sort
of counterparts) seems to be detected in nearby clusters, and prefer a
scenario where the high-z blue galaxies have evolved into another kind
of galaxy type.

While such a population cannot possibly transform into early--type
galaxies which are old, both locally (Bower, Lucey \& Ellis 1992,
Andreon 2003) and up to $z=1$ (Stanford, Eisenhardt, \& Dickinson
1998, Kodama et al. 1998, Andreon, Davoust, \& Heim 1997; Ellis et
al. 1997, but see van Dokkum \& Franx 2001 for a different opinion),
S0 galaxies could provide a destiny (but see Ellis et al. 1997;
Andreon 1998b; Jones, Smail, \& Couch 2000 for a different
opinion). The relative fractions
of spirals and S0s observed in clusters at different redshifts 
(Dressler et al. 1997) seem
to support such morphological transformations (but see Andreon
1998b; Lubin et al. 1998 for a different opinion). 

However, Allington-Smith et al. (1993) argue that galaxies in groups
do not evolve (except passively), at least over the redshift interval
$0 < z < 0.5$, and suggest that the BO effect should be interpreted as
an evidence of the important role played by the cluster environment:
evolution is strong in rich clusters and negligible (because
inefficient) in poor environments. 

Whether the BO effect has been confirmed or not is unclear:
by studying clusters
at very similar redshifts, Smail et al. (1998) and Pimbblet
et al. (2002) do not find an increase of the blue fraction
with redshift, although 
Margoniner \& de Carvalho (2000) and Margoniner et al. (2001) do.
Fairley et al. (2002) observed clusters at higher redshift 
and did not find any signature of a BO effect.
Kodama \& Bower (2001) and Ellingson et al. (2001) reach opposite
conclusions on the existence of an excess of blue galaxies in the
cluster core, the former paper using a subsample of the data used
in the latter. 
Neither the amplitude, when the BO effect is detected, is the same in 
different works: Rakos \& Schombert (1995) tend to find a larger
blue fraction (at a fixed redshift) than Butcher \& Oemler (1984).

The existence of the BO effect has also been criticised or, simply,
not found when expected to show up markedly: a high blue fraction at
high redshift has not been confirmed by van Dokkum et al. (2000) 
for the X-ray cluster MS 1054-03 at $z=0.83$.  Apart from
all criticisms raised before 1984 and addressed in Butcher \& Oemler
(1984), Kron (1994) claimed that all the "high" redshift clusters
known at the time were somewhat extreme in their properties, and this
was precisely what had allowed them to be detected. Observations of
four clusters at $z \sim 0.4$ led Oemler, Dressler, \& Butcher (1997)
to suggest that clusters at that redshift are more exceptional
objects than present-day clusters, and are actually being observed
both in the act of hosting several galaxy-galaxy mergers and
interactions, as well as growing by merger of smaller clumps, in
agreement with a hierarchical growth of structures as described, for
example, by Kauffmann (1995). The higher infall rate in the past would
also favour higher blue fractions in distant clusters.
Andreon et al. (1997) have made a detailed comparison of the
properties of galaxies in the nearby Coma cluster and cl
0939$+$47 at $z=0.41$. They found that the spiral population of these two
clusters appears too different in spatial, colour, and surface
brightness distributions to be the same galaxy population observed at
two different epochs. The Coma cluster is therefore unlikely to be
representative of an advanced evolutionary stage of cl 0939$+$47, and so
any comparison between the blue fraction of the two systems may be
delusive. Andreon \& Ettori (1999) raised two more concerns: the BO
sample does not form a homogeneous sample of clusters over the studied
redshift range; furthermore, optical selection of clusters is prone to
produce a biased - hence inadequate - sample for studies on evolution
since, at larger redshifts, it naturally favours the inclusion in the
sample of clusters with a significant blue fraction. This argument is
also presented by de Propris et al. (2003), who also argue that the BO
effect is due to the optical selection of the galaxies: low mass
galaxies with active star formation have their optical colour boosted,
and these galaxies increase the cluster blue fraction.

\medskip 

This paper has two aims: to extend the measurement of the blue
fraction to a redshift range largely not probed yet, and to critically
review and discuss the analyses performed thus far by various authors
in the literature.  We studied three clusters selected among  
best detected and possibly at high redshift
cluster candidates listed in Lobo et al. (2000) and detected
in the EIS data set (Nonino et al. 1999): cl2249--3958, cl2245--3954
and cl0048--2942. For each one of these three clusters we have redshift
information available, confirming the presence of a galaxy overdensity
at redshift 0.71, 0.66, and 0.64 respectively (Serote Roos et al. 2001).

In section \S 2 we present the data for the three clusters and for the
control fields, and in \S 3 we detail the step by step description of
the determination of the blue fraction of the three clusters. Results
are presented in \S 4 where we also critically re--examine previous
studies on the BO effect. Finally, we summarize the results and
conclude in \S 5.

\smallskip 

We adopt $\Omega_\Lambda=0.7$, $\Omega_m=0.3$ and $H_0=50$ km s$^{-1}$
Mpc$^{-1}$. The $H_0$ value was chosen for consistency with previous
works, but is largely irrelevant for the results of this paper
because it cancels in the comparisons (eg. in the difference of
distance modulii or in the ratio of metric diameters). 

\begin{figure*}
\centerline{%
\psfig{figure=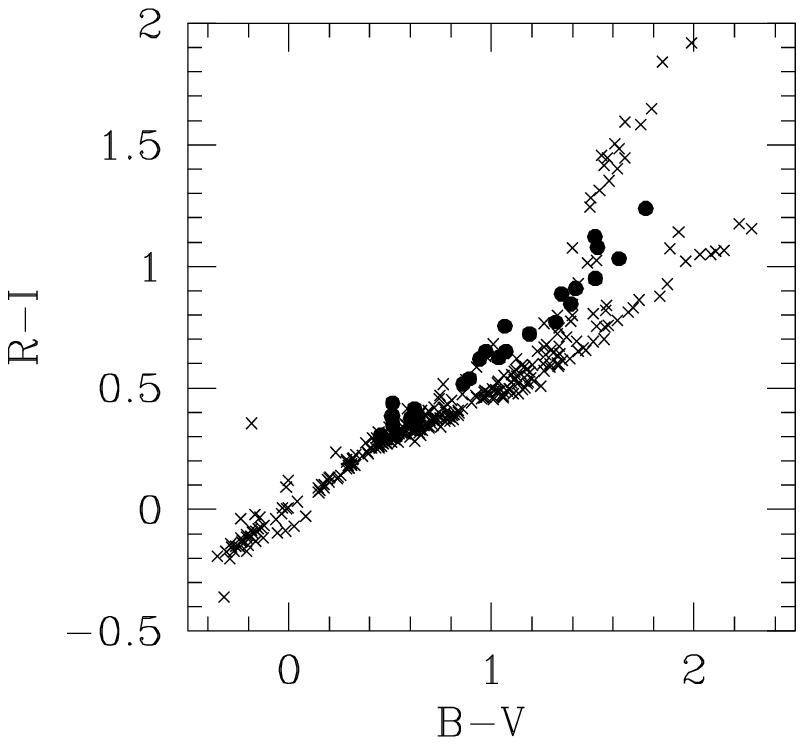,width=8truecm}%
\psfig{figure=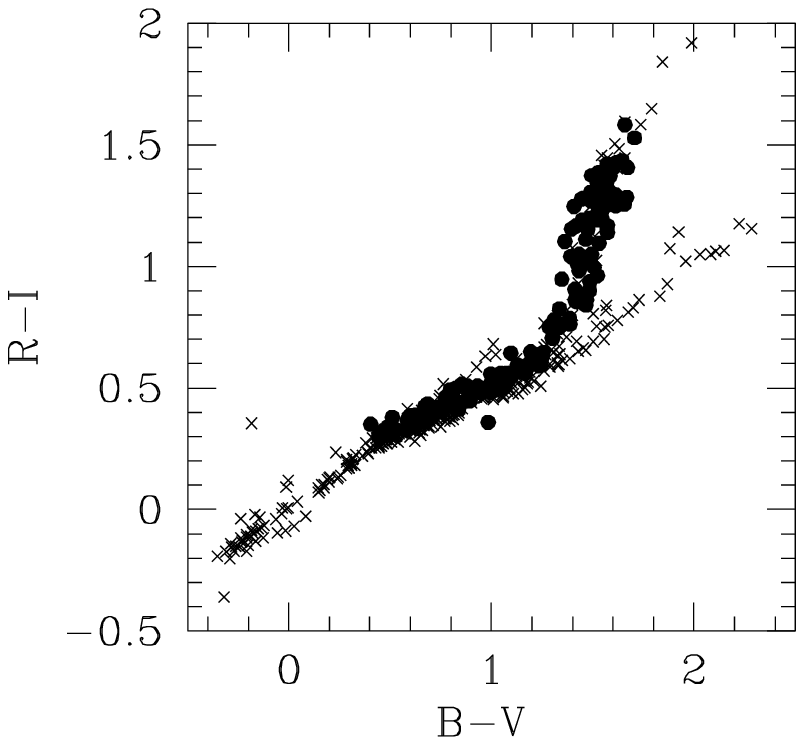,width=8truecm}}
\caption[h]{Colour--colour diagrams for the stars. 
Crosses indicate their colours as given by Landolt 
(1992) whereas circles refer to the measurements we performed in our
fields. The right panel refers to our control field, whereas the left
panel concerns the three cluster fields. Note the good agreement
between the expected and observed star loci in all cases. In the
left panel there are 26 stars measured in our cluster fields, most of
them falling in the crowded part of the diagram and hence not easily
visible in this plot.}
\end{figure*}

\section{The data}

Observations of the three clusters were performed
at the 3.6 m ESO telescope at La Silla with the EFOSC2 camera, 
in the Bessel--Cousins $B,V,R$
and $I$ filters and with the Loral \#40 CCD. Clusters cl2249--3958 and
cl2245--3954 were observed in the first run (25-26-27 October 1998;
run ID 62.O-0806) whereas cl0048--2942 was targeted during our second
run (14-15-16 August 1999; run ID 63.O-0689). The CCD field of view
provided images covering $\sim 5 \times 5$ arcmin$^2$ but a slightly
different strategy (pixel binning and exposure time per filter) was
adopted in each run: these and other details on the data are provided
in Table \ref{tab:data}. At $z=0.7$, 3 Mpc subtends a 5 arcmin angle.
$r_{200}$ corresponds to $\sim0.8$ Mpc, or $1.4$ arcmin, for a cluster 
at $z=0.7$ and having a velocity dispersion of 500 km/s.

\begin{figure}
\psfig{figure=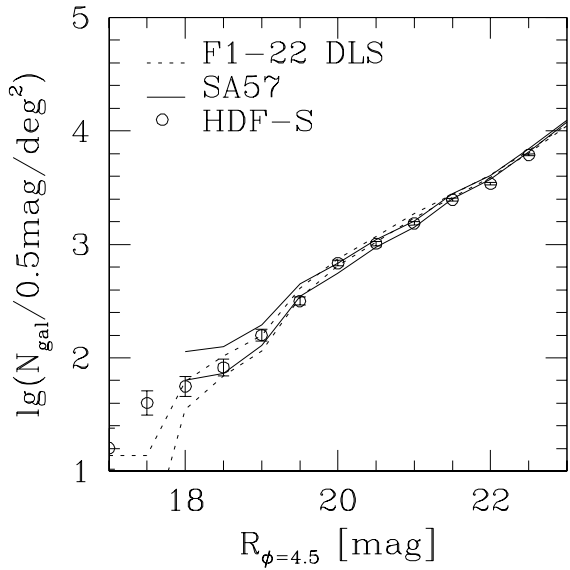,width=8truecm}
\caption[h]{
Galaxy counts in the adopted control field (circles) and in the
direction of two other control fields: the Selected Area 57 at the North
Galactic Pole, and the F11--22 area of the Deep Lens Survey. There is no
evidence for a discrepancy in galaxy counts in the HDF--S with respect
to the other control fields. The areas of the three fields are,
roughly, 0.50, 0.18 and 0.34 deg$^2$ for the HDF--S, SA57 and F11--22,
respectively. Magnitudes were measured in an aperture of 4.5 arcsec.}
\end{figure}

\begin{figure}
\psfig{figure=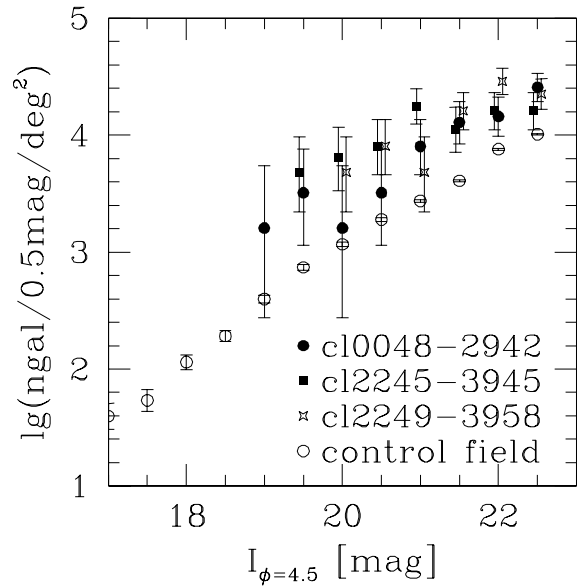,width=8truecm}
\caption[h]{Galaxy counts in the control field (open circles) and in 
the direction of the three clusters. There is a clear excess of
galaxies in the line of sight of each cluster with respect to the
background galaxy counts. }
\end{figure}

Serote Roos et al. (2004) present
a detailed description of the spectroscopic data and analysis. Shortly,
targets for spectroscopy were selected from the photometric catalogueues
using priority criteria based on colours and compactness. Spectroscopic
observations were performed at VLT with FORS1 and FORS-2 in 1999 and 2000,
at a resolution of about 500, and with a typical exposure time of
1 hour. We have at least 22, 11 and 7 consistent
redshifts for galaxies in the fields of cl0048--2942, cl2245--3954 and
cl2249--3958, respectively, supporting the existence of a
gravitationally bound system at $z \sim 0.64, 0.66$ and $0.71$ in each
case (see Serote Roos et al. 2001 for an early report). 

Photometric data from the 3.6 m have been bias--subtracted,
flat--fielded and fringe--corrected (when needed in the redder filters
$R$ and $I$) using basic IRAF routines. 
Then, cosmic rays were detected and flagged by using task
FILTER/COSMIC in Midas. Pairs of images were aligned and then combined
by means of {\em imcombine} in IRAF. In this step we made full use of
the masks that flagged both permanent defective pixels and cosmic
rays, and took into account differences in airmass between the
combined images. Finally, we kept only the sky region fully exposed in
common to all filters (we note that we had applied small shifts at the
telescope, of the order of 5 pixels, for optimizing the sampling
especially for regions covered by defective areas of the CCD; 
this produced slightly different fields of view for
each pointing).

Objects have been detected by SExtractor v2 (Bertin \& Arnouts 1996), in double
image mode and using the $I$ band image for detection, in order to assure that
colours are available for all $I$ band detected galaxies and that the resulting
catalogue is actually complete in $I$, whatever the luminosity of the objects in
the other filters is. For the star/galaxy classification we used the SExtractor
Class\_Star index and we discarded only objects more
compact than Class\_Star=0.95 in the $I$ band and brighter than $I=21$ mag. 
Fainter stars are statistically removed by using the control field,
in the same way as for foreground and background
galaxies (see, for example, Andreon \&
Cuillandre 2002). This way we avoid rejecting compact galaxies
that could be misclassified as stars because their compactness.
Magnitudes have been measured  within a 4.5 arcsec aperture.

We neglected galactic absorption, that accounts for 0.01 mag at most
in $R-I$ (Schlegel, Finkbeiner, \& Davis 1998), the colour
used in our BO analysis.

Data were calibrated by observing several Landolt (1992) standard
stars. The large number of standard stars observed during the August
99 run allowed us to compute colour terms for our system. These turned
out to be very small (of the order of 0.03 at most per unit colour). 
We checked our photometric calibrations by comparing the star
loci, in the colour--colour plane, of Landolt stars and of stars in our
field of view (as, for example, in Puddu, Andreon, Longo et al.
(2001)), and in the colour {\it minus} colour {\em versus} colour plane, in order
to remove the strong correlation between colours and emphasize
systematic errors. The left panel of Figure 1 shows the good
agreement between the expected and observed star loci 
for the stars in our cluster
fields: our star sequence falls on top of the Landolt one.
All colours, therefore,
do not present any problem at the 0.01-0.02 mag level.  All nights
were photometric, with residual zero--point variations of 0.02 mag for
$V, R, I$ and 0.04 mag for $B$.

The resulting catalogue turns out to be complete to $m_I\sim22.5-23$
(see Table 1),
which corresponds to evolved $M_V\sim-20.0 $ to $-19.5$ mag 
(see Section 3.1 for the $m_I$ to $M_V$ conversion). 
Completeness was estimated as in Garilli, Maccagni \& Andreon
(1999), Andreon et al. (2000) and Andreon \& Cuillandre (2002) by
looking at the magnitude of the brightest galaxies having the lowest
detected central surface brightness.

As control field we used images of the Hubble Deep Field South
(HDF-S), retrieving the data from the Goddard group (see
http://hires.gsfc.nasa.gov/$\sim$research/hdfs-btc/ and Palunas et al.
2002), as released on April 15th 2002. These data were taken with the
Big Throughput Camera (Wittman et al.\ 1998) on the Blanco 4 m
telescope at CTIO and calibrated in the Landolt photometric system by
observing Landolt (1992) standard stars, as for our program images.
The camera is a $4k\times 4k$ mosaic with large CCD gaps. The area
surveyed is very large, covering 0.5 square degrees, large enough to
get rid of the cosmic variation of galaxy counts.
From that area we excluded a few regions, where obvious clusters are
located, with no considerable change to the area surveyed. We produced
the corresponding galaxy catalogue exactly as previously done for our
cluster fields.  The right panel of Figure 1 shows the locus of stars
in the colour-colour plane, comparing values taken directly from the
Landolt catalogue, and those measured by us in the control
field image. The agreement between the two loci is good,
showing that the control field observations are indeed in the same
photometric system of our cluster observations.

Two secondary control fields were used, in order to check that the
$\sim$ 0.5 deg$^2$ area of the HDF--S is a typical sky region devoid
of any particular large scale structure. The first one is the F11-22
area of the Deep Lens Survey (public images are at
http://dls.bell-labs.com/Publicdata/index.html), observed in BVRI with
the Mosaic camera at the 4 m KPNO telescope. The field of view of the
image is about 1/3 square degrees. The other secondary control field
is the Selected Area 57 at the North galactic pole, whose images are
presented in Andreon \& Cuillandre (2002). Here we use only the
$R$--band images, that were taken with the UH8k camera (Luppino et
al., 1994) at the CFHT. The field of view of that image is
about 0.2 deg$^2$. Both the UH8k and the Mosaic camera are $8k\times
8k$ devices formed by tightly packed CCDs.

Again, for these two additional control fields we produced the
corresponding galaxy catalogues with SExtractor, exactly with the same
general settings as applied to the cluster images and the HDF-S
control field.

The three control fields are well apart in the sky: the HDF-S is at
$22^h34^m~-60^o37'$, the SA57 at $13^h09^m~+29^o09'$ and the F11-22
field at $0^h53^m~+12^o35'$, and therefore sample three very different
lines of sight.  Figure 2 shows the galaxy counts for the three
control fields in the $R$ band, by adopting a 4.5 arcsec aperture
magnitude. There is a good agreement among the counts measured in the
three different lines of sight, showing that none of the three
pointings is peculiar in galaxy density. All our three control fields
are deeper than programme cluster images, and hence can be used
in this work.

Figure 3 shows the galaxy counts for the central regions of the
cluster pointings and for the HDF-S control field in the $I$ band, for
the same aperture magnitude. There is a clear, and reassuring, excess
of galaxies in the clusters lines of sight.

\begin{figure*}
\centerline{%
\psfig{figure=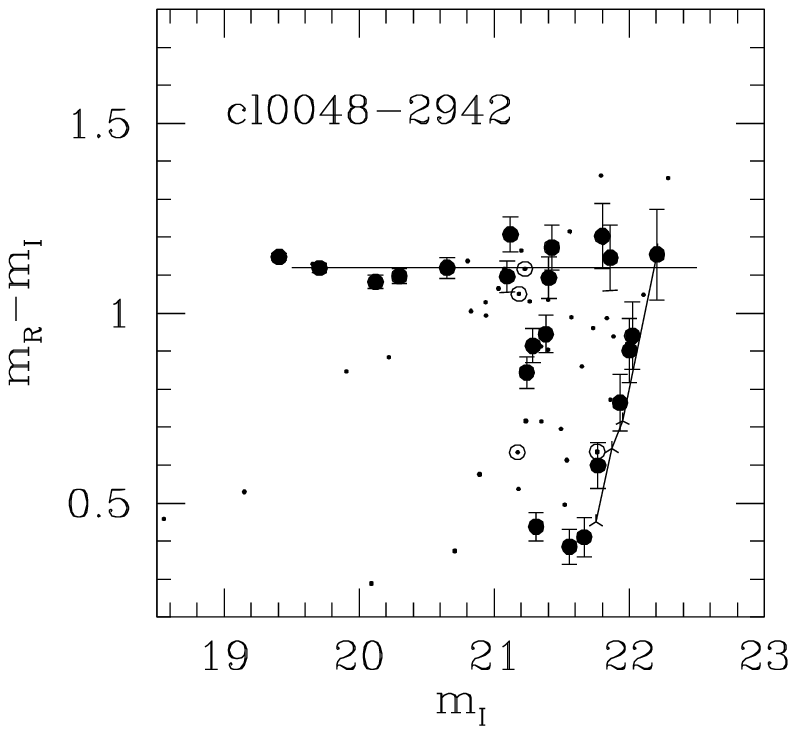,width=6truecm}%
\psfig{figure=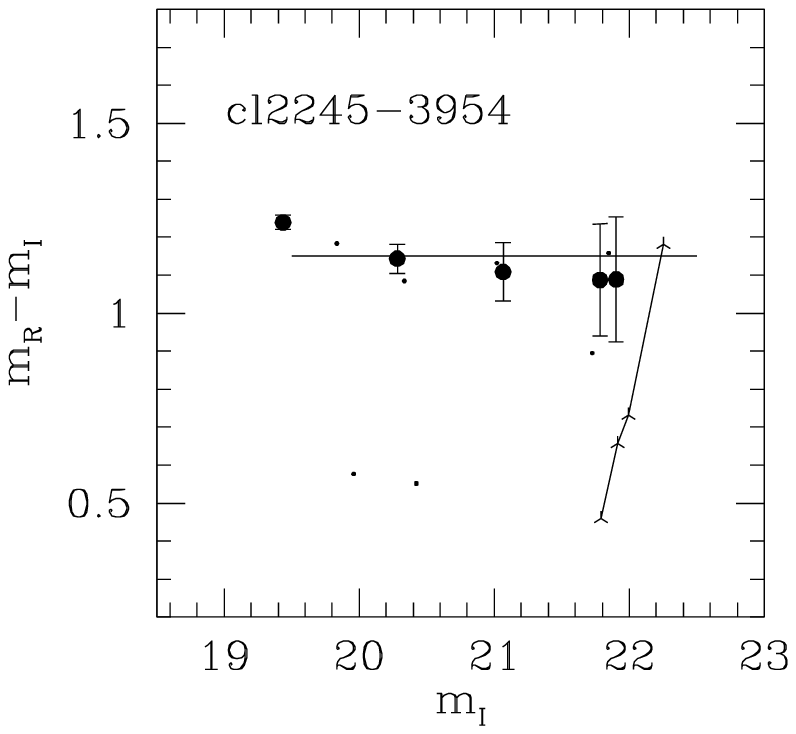,width=6truecm}%
\psfig{figure=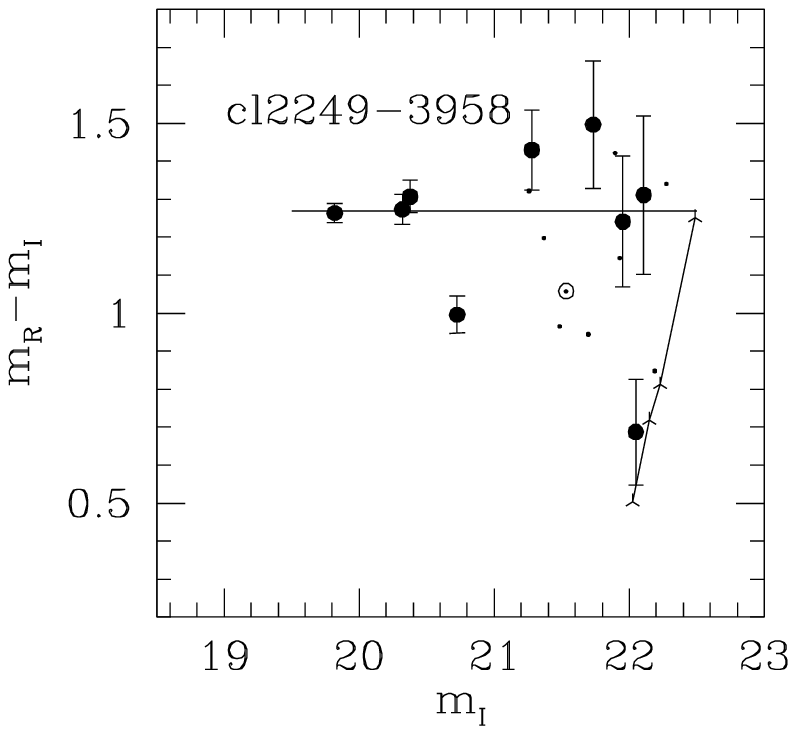,width=6truecm}}
\caption[h]{Colour--magnitude diagrams for the three cluster fields 
considering all galaxies within $R_{30}$. The straight
horizontal line marks the colour of the red sequence, whereas the slanted
line marks the evolved $M_V=-20$ mag cut. Filled symbols are statistical members,
whereas open points are statistical interlopers. Small points mark galaxies
falling inside $2R_{30}$.}
\end{figure*}

\section{Determination of the blue fraction}

BO define the fraction of blue galaxies, $f_b$, in the cluster as the
fraction of galaxies bluer by at least 0.2 mag, in the $B-V$
rest-frame, than early--type galaxies at the cluster redshift. The
galaxies have to be counted down to an absolute magnitude of $M_V=-20$
(for H$_0=50$ km s$^{-1}$ Mpc$^{-1}$), within the radius $R_{30}$ that
encompasses 30\% of the cluster galaxies. Moreover, galaxies in the
background or foreground of the cluster have to be previously removed,
for example by statistical subtraction.

The {\it actual} BO limiting magnitude is, at the BO high redshift end,
brighter than $M_V=-20$ mag (see de Propris et al. 2003). 
A brighter limiting magnitude at higher redshift
is the correct choice if one wants
to track the same population of galaxies at different redshifts,
because of average luminosity evolution experienced by
galaxies.  Therefore, we adopted an evolving $M_V=-20$ mag limit, as 
{\it actually} adopted by BO. An evolving limiting magnitude has also
been adopted by de Propris et al. (2003) and Ellingson et al. (2001)
in their BO--style studies.

According to measurements by Lin et al. (1999) of the
evolution of $M^*$ in the $R$ band in $0.12<z<0.55$, we expect to have
approximately 1 mag brightening in the $V$ band galaxy luminosities at
the redshift of our clusters for our adopted cosmology, in
agreement with Bruzual \& Charlot (1993) models. Therefore, we
adopt a 1 mag brighter limiting magnitude than the non evolving cut.

\begin{table*}
\caption{Cluster characteristics}
\begin{tabular}{ccccccccc}
\hline
Cluster & z & $N_z$ & $R_{30}$ & Red Sequence & $\Delta(m_R-m_I)$ & $C$ & $N_{red}$ &
$N_{all}$\\
name &  &  & [arcmin] & $m_R-m_I$ [mag] & [mag] & & &\\
\hline
 cl0048-2942  & 0.64 & 22 & 0.8  & 1.12 &  0.26 & 0.4 & 40 & 50\\
 cl2249-3958  & 0.71 & 11 & 0.2  & 1.27 &  0.26 & 0.5 & 30 & $\sim$ 40 (see Notes) \\
 cl2245-3954  & 0.66 & 7  & 0.2  & 1.15 &  0.26 & 0.4 & 15 & 25 \\
\hline
\end{tabular}
\medskip
\break \noindent
Redshifts are from Serote Roos et al. (2001). \hfill \break
$N_{red}$ is the asymptotic number of member galaxies brighter 
than $m_I=22.5$ mag (which corresponds roughly to the evolved $M_V=-20$ mag) integrated 
over the whole ``growth curve" that are red according to the BO definition. 
\hfill \break \noindent
$N_{all}$ refers to the same calculation but without applying 
any colour cut. We note that this value could not be computed directly 
for cl2249--3958 because of a contaminating population in 
the outskirts of this cluster; it was therefore computed assuming
the same global blue fraction of the cluster cl0048-2942, that has a similar
blue fraction within $R_{30}$ and $2\times R_{30}$ (see section 
\ref{sec:res}). \hfill
\label{tab:charact}
\end{table*}

\subsection{Colour and magnitude cuts and K-corrections}

In order to measure the fraction of blue galaxies we must compute the
colour cut in the observer's frame, $\Delta(m_R-m_I)$, that corresponds to
the rest-frame $\Delta(B-V)=0.2$ set by BO. To this end, we note that
the $(B-V)$ colour difference at zero redshift between an E and an Sbc
is 0.34 mag (Frei \& Gunn 1994), when the spectral energy
distributions are taken from the Coleman, Wu, and Weedman (1980)
templates. The colour cut in the observer frame $\Delta(m_R-m_I)$
corresponding to the original BO definition will therefore be
$\Delta(B-V)=0.2/0.34$ times the colour difference between an E and an
Sbc at the cluster redshift according to the same models.
For our three clusters, we get $\Delta(m_R-m_I)=0.26$ mag. 

K--corrections, needed to perform the above calculation, have been
computed according to Weinberg (1972) using the response curve of our
filters (listed in the Midas environment) together with the quantum
efficiency of the Loral CCD. We adopted as reference galaxy spectra
those of Coleman, Wu \& Weedman (1980).  We checked our code verifying
that we get the same K--corrections as previous work in the
literature (Frei \& Gunn 1994). For synthetic colour measurements we
use the Vega spectrum that is included in the GISSEL96 distribution
(Bruzual \& Charlot, 1993). 


For each cluster, the median colour of the three brightest galaxies
in the $(m_R-m_I)$ {\em versus} $m_I$ colour--magnitude relation was assumed as
the typical colour of early--type galaxies at the cluster redshift (and can
be found in the fifth column of Table 2). Puddu,
Andreon, Longo et al. (2001) show that such a measure tracks the colour
of the red sequence for their clusters, that spread through
$0<z<0.35$. Our observed colours well agree with those expected for 
early--type passive evolving galaxies at the redshift of our clusters.

In order to compute the corresponding $M_V$ absolute mag cut, we
should take into account the fact that the match between the $I$
filter and the rest-frame $V$ is only approximate. This implies that
galaxies of identical $M_V$ magnitudes have slightly different $m_I$
magnitudes, with differences correlated to colour. Therefore, the cut
$M_V=-20$ of BO's definition becomes, when translated to a cut in
observed $m_I$ magnitudes, slightly colour dependent.  Adopting
K-corrections in the $R$ and $I$ filters  
and rest-frame $(R-I)$ colours for the spectral
templates listed in Coleman, Wu \& Weedman (1980) we obtain
the slightly slanted (almost vertical) line in Figure 4. The galaxies
to be considered for the BO effect are those to the left of this line
(and for this reason fainter galaxies are not plotted).  
This cut has been applied to cluster and control field samples.

The slope of the colour--magnitude relation and its impact on the
determination of the fraction of blue galaxies is negligible (and
neglected).

\begin{figure*}
\centerline{%
\psfig{figure=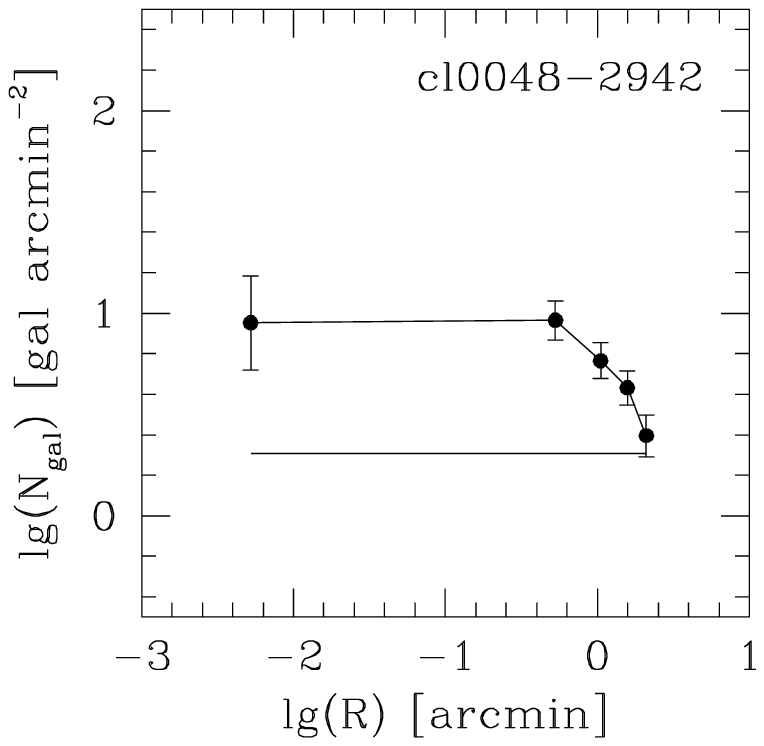,width=6truecm}%
\psfig{figure=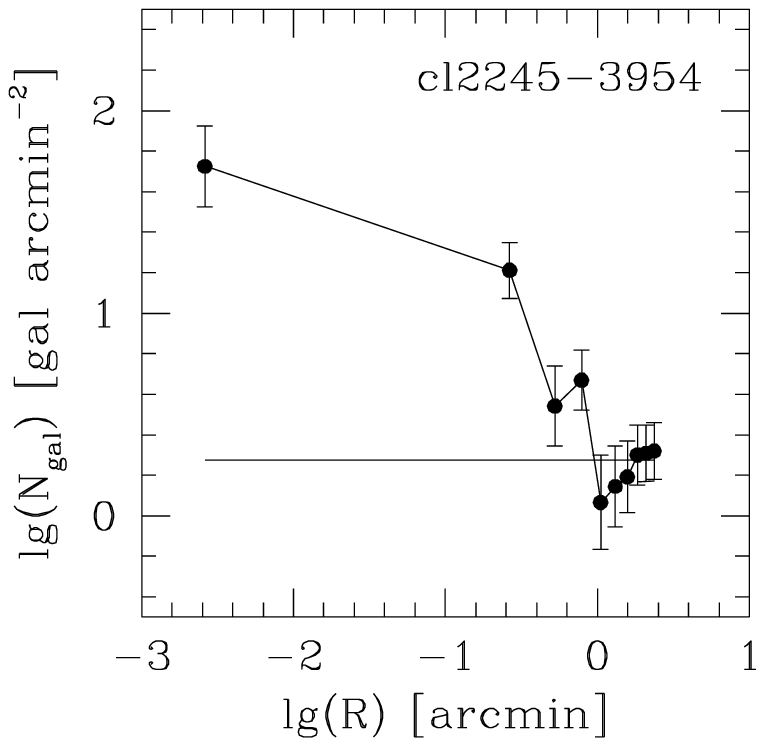,width=6truecm}%
\psfig{figure=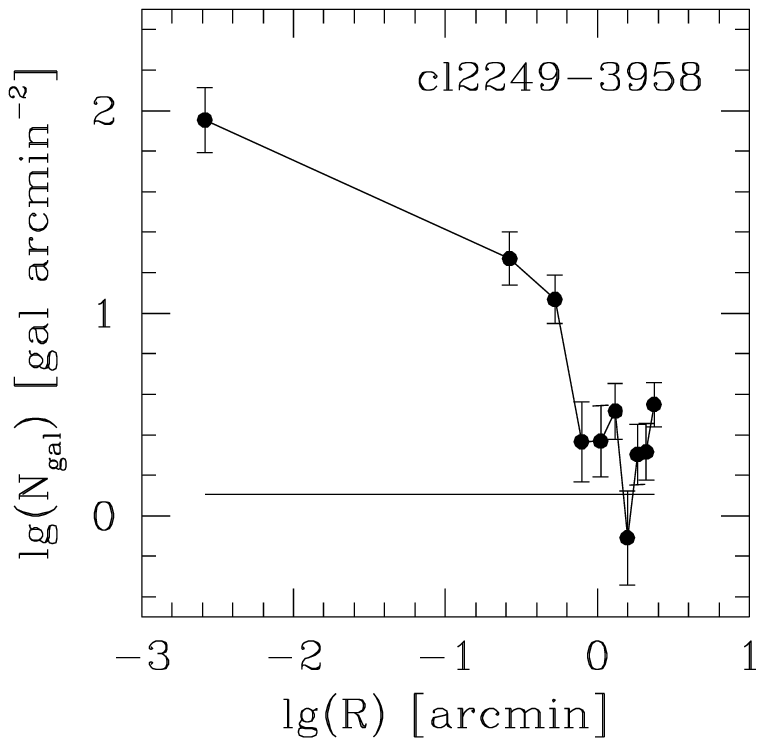,width=6truecm}}
\caption[h]{Radial profiles for the three clusters considering all red galaxies
(according to the BO definition) brighter than I=22.5 mag and
previously to any background subtraction. The expected
background, independently measured on the control field, is shown by
the horizontal line. It slightly differs, from cluster to cluster, because
of differences in the considered colour range.}
\end{figure*}

\subsection{Cluster radii and $f_b$'s}

In order to derive the fraction of blue galaxies in our clusters we
first need to estimate $R_{30}$, the radius including 30\% of the
cluster members.  We computed the radial profile, $N(r)$, of our three
clusters by counting galaxies in circular rings centred on the brightest
cluster galaxy (we did take into account that
our rectangular field of view truncates the outermost rings) and by
statistically subtracting the background galaxy density using our
control field. We used rings of increasing width outwards, in order
to keep the S/N almost constant.

In order to enhance the contrast between the cluster and
the background when counting $N(r)$, we considered only galaxies brighter
than $m_I=22.5$ mag and  within 0.26 mag from the colour--magnitude 
relation (i.e. red galaxies
according to the BO definition). Needless to say, this colour selection
was applied only in this step and consequences on $R_{30}$ are quantified
below.  The radial
profiles, $N(r)$, for the three clusters are shown in Figure 5, together with
the expected background counts obtained from the control field for each
cluster area for the same magnitude and colour cuts.
In a similar way, we computed the integrated radial
profiles or ``growth curves", $N(<r)$. 

The concentration index $C$ (Butcher \& Oemler 1984) is defined by:

$C=\log(R_{60}/R_{20})$

\noindent
where $R_{60}$ and $R_{20}$ are the radii that include 60 and 20\% of
the cluster members. The three clusters have values that classify them
as "compact" according to BO ($C = 0.4-0.5$, see Table
\ref{tab:charact}), i.e. that qualify them as appropriate for the BO
measure of the fraction of blue galaxies.

BO, working with nearer clusters, could avoid applying the colour selection
we adopted for our clusters. The colour cut is especially required for
cl2249-3958 because of a blue contaminating population at large
cluster--centric radii, that prevents the growth curve from converging
to a constant value. Spectroscopic observations (Serote Roos et al.
2001) confirm such a contamination.  We do stress that, for the other two clusters,
$R_{30}$ remains unchanged, within 10\%, when derived using the whole
galaxy population or the red population only, therefore confirming that
having chosen only red galaxies does not bias our results.

\begin{figure}
\centerline{%
\psfig{figure=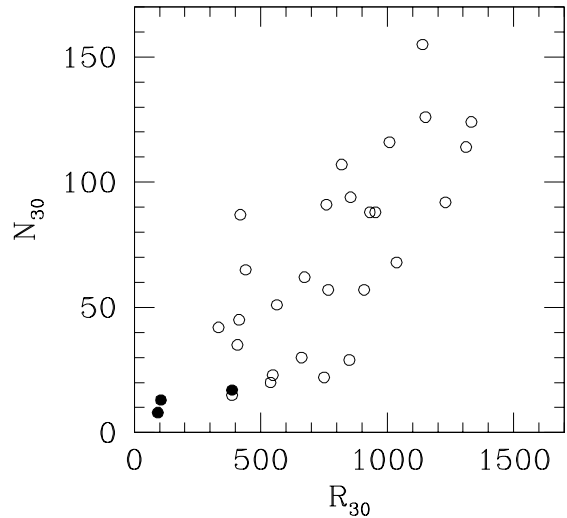,width=8truecm}}
\caption[h]{Relation between $N_{30}$ and $R_{30}$ (in Kpc) for the BO clusters 
(open circles) and our sample (filled circles). Errors on $R_{30}$ for
our sample are estimated to be of the order of 10 to 20 \%.}
\end{figure}

All three of our clusters have small estimated values of $N_{30}$  
and $R_{30}$ relatively to the BO sample (see Table 2 and Figure 6).
Our smaller $R_{30}$ radii are not 
due to an error in the background subtraction
because the contrast between cluster and
field is quite high at $R_{30}$. Furthermore, the background is not a
free parameter: it was estimated on a wide sky area ($\sim$ 0.5
deg$^2$ wide), which renders the average value (and its variance)
very well determined.
We can exclude
that our main control field is ``anomalous'' (too dense or too
sparse), or a possible
mistake in the photometric calibrations, as already 
discussed in Section 2. We are hence confident that 
our estimate of $R_{30}$ is reliable.

\medskip
We can now proceed to measure $f_b$. Some of the galaxies in each cluster
line of sight will be interlopers, i.e. not physical members of the
cluster but just objects projected  along the line of sight, and we remove
them statistically in exactly the same way as BO did.

We note that the cluster mass might alter, through lensing, the
luminosity of the background population, therefore increasing the
galaxy counts locally. This would artificially lead to an overestimate
of the cluster counts. The change in the galaxy counts due to the lensing 
is a power of $2.5 \times (0.4-\alpha)$ (Bernstein et al. 1995) where
$\alpha$ is the slope of the galaxy counts. In the $I$ band, i.e. in
the filter we used for selection, we found a slope of 0.40 between
19.0 and 22.5 mag, the magnitude range in which we are interested in. 
Therefore, lensing has a very small, if any, effect upon the
background counts.  

We used the control field to estimate the
expected number of interlopers in $R_{30}$ within our magnitude cut,
their Poissonian fluctuations, and the blue fraction in our clusters. 
We repeated this operation 100 times, each time performing a different
extraction from the control field. We then computed the median blue
fraction and the scatter around the median. Results are shown in 
Table \ref{tab:bo}.

The scatter computed thus far does not take into account the cosmic variance,
i.e. the variance, in excess to Poissonian fluctuations, of
galaxy counts. We remind the reader that no matter
how well the mean background is determined, what limits the accuracy
of the background subtraction is the background variance on the
spatial scale where $R_{30}$ is measured. We divided
the area of the control field in cells, each cell having
an area equal to $\pi R_{30}^2$, and counted the frequency with
which we observe $N$ galaxies (with $0<N<\infty$), therefore
deriving the background variance on  $R_{30}$ scale. 
For galaxies brighter than $m_I=23$ mag we observed a variance that is
47\% (12 \%) larger than expected for a Poissonian distribution
for $R_{30}=0.8$  $(0.2)$ arcmin. Therefore, 
the Poissonian term listed in the last column of Table 3
should be multiplied by 1.47 or 1.12, depending on $R_{30}$.

\begin{table}
\caption{Cluster blue fractions, $f_b$, measured within radius R and down to the limiting magnitude $M_V$.}
\begin{tabular}{lclll}
\hline
Cluster name & R  & $M_V^e$ & $f_b$ & $\sigma(f_b)$ \\
\hline
cl0048-2942 & $1\times R_{30}$ & -20.0 & 0.29 & 0.05 \\
            & $1\times R_{30}$ & -19.5 & 0.33 & 0.05 \\
            & $2\times R_{30}$ & -20.0 & 0.25 & 0.06 \\
            & $2\times R_{30}$ & -19.5 & 0.29 & 0.05 \\	   
cl2249-3958 & $1\times R_{30}$ & -20.0 & 0.20 & 0.04 \\
            & $1\times R_{30}$ & -19.5 & 0.25 & 0.03 \\
            & $2\times R_{30}$ & -20.0 & 0.22 & 0.05 \\
            & $2\times R_{30}$ & -19.5 & 0.26 & 0.05 \\    
cl2245-3954 & $1\times R_{30}$ & -20.0 & 0.00 & 0.00 \\
     	    & $2\times R_{30}$ & -20.0 & 0.17 & 0.05 \\
\hline
\end{tabular}
\label{tab:bo}
\medskip
\break \noindent
The quoted $\sigma$ is half the interquartile range. For a Gaussian
distribution the dispersion is 1.47 times the half interquartile
range. \hfill \break
Errors listed in this Table are as in literature, and
do not take into account our discussion in Sect 4.5.\hfill \break
 $^e$ Magnitudes corrected for evolution. 
\end{table}

\section{Results \& Discussion}\label{sec:res}

\begin{figure}
\centerline{%
\psfig{figure=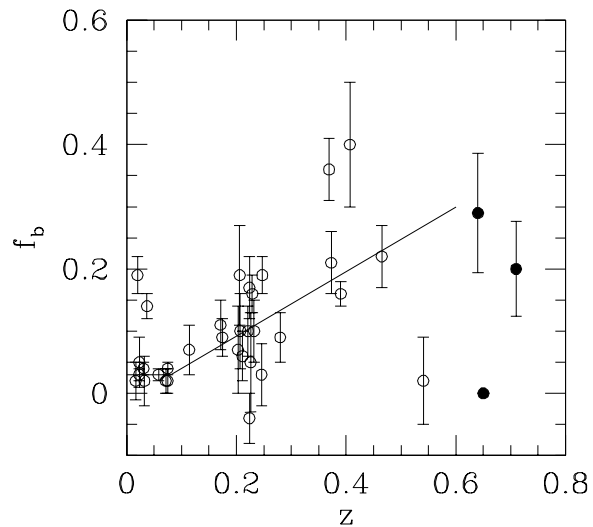,width=8truecm}}
\caption[h]{Blue fraction as a function of redshift. Open circles mark 
the BO sample clusters with the respective error bars as published by BO;
filled circles indicate our clusters. Error bars do not include the
error coming from the sample representativity (see Sect. 4.5). The spline
is the Butcher \& Oemler (1984) eye fit to the data.}
\end{figure}

Table \ref{tab:charact} summarises the measured cluster
characteristics: their redshift, $R_{30}$, the number of members
with known redshift, $N_z$,
the colour of the red
sequence, the adopted colour cut, the concentration index and the (asymptotic)
number of member galaxies brighter than $m_I=22.5$ mag. The latter
quantity is computed twice: for red galaxies ($N_{red}$, 8th column) and without
any colour selection ($N_{all}$, last column). $N_{all }$ is 
comparable to the cluster richness as measured by Abell (1958):
the magnitude range in which galaxies are counted is very
similar and our asymptotic measurement of $N_{all}$ is equivalent to
the 3 Mpc radius adopted by Abell (1958) in order to encompass the
whole cluster.  
The three clusters have, therefore, $R=0$ to $1$, $R$ being the standard
(Abell 1958) richness.

In the BO sample the cluster richness increases with redshift (Andreon
\& Ettori 1999): the highest redshift clusters are of
richness $R=3$ or $R=4$, being by far the richest of all the sample,
a result of a bias in the cluster sample available at that time
(see also Kron 1994). Our three clusters are extracted from the EIS survey,
which covers a small sky area. As a consequence the probability
of getting a very rich cluster is low and this is why our sample contains
common and lower richness clusters, which turns out to be comparable in
richness to the low redshift BO sample.

Table \ref{tab:bo} shows the blue fraction, $f_b$, of our three clusters
computed at different radii for galaxies brighter than two evolved
limiting magnitudes. This Table also lists Poissonian
errors on the blue fraction, $\sigma(f_b)$.

While postponing to the next sections a thorough discussion of the values 
listed in Table 3,
we compare in Figure 7 the values derived for the BO sample and for our three 
clusters. The BO {\it extrapolated} value of the blue fraction is
around 0.35 at the mean redshift of our three clusters.
cl0048-2942 has a blue fraction  
compatible with the extrapolation of the BO
linear trend.  The two other clusters have lower blue fractions.
Our data, therefore, do not show any strong evidence for the presence
of an increasing $f_b$ with look--back time, in agreement with
the approximately constant 
fraction of blue galaxies within $0.5 r_{200}$ 
found by Ellingson et al. (2001).
Before drawing any final conclusion from this plot it is instructive to take
a deeper look at the analysis performed by other authors in
the literature because different, and sometimes contradictory, results
have been obtained on the BO effect.

\subsection{Luminosity dependence}

In this paragraph we will examine how $f_b$ changes as a function of 
the adopted luminosity cut--off.
At low redshift luminosity functions of different morphological types
have different shapes (e.g. Bingelli, Sandage \& Tammann 1988; Andreon
1998a) and therefore it is
likely that the luminosity function of galaxies of different colours
differs and that the fraction of blue galaxies will depend on the
luminosity cut--off. On the other hand, BO have shown that as long as
this cut--off value is in the $-22<M_V<-20$ mag range, the blue
fraction does not depend on the exact value of the limiting magnitude
for four of their clusters. A similar
result was found by Kodama \& Bower (2001) for 6 EMSS
clusters and by Ellingson et al. (2001) on a sample of 
15 clusters. Fairley et al. (2002)
observed a more complex situation: in 5 out of 8 clusters the blue
fraction was identical when a $M_V=-21$ mag or $M_V=-20$ mag cut--off was
adopted, whereas for the three remaining cases $f_b$ was lower by 0.1 (a
1 $\sigma$ effect) when the brighter cut--off was adopted.  For two of our
clusters, cl0048-2942 and cl2249-3958, we can make a measure of the BO
effect both at evolved $M_V=-20.0$ and $M_V=-19.5$ limiting magnitudes: the
$f_b$ of these two clusters remains constant within 1 $\sigma$ (see
Table \ref{tab:bo}).  For the third cluster, cl2245-3954, we cannot perform
such a comparison because data are not deep enough.

Therefore, $f_b$ does not seem to depend critically on the adopted luminosity
cut--off.

\subsection{Richness, radial profile and shortcomings of a unique metric aperture}

In this paragraph we will examine how $f_b$ changes as a function of the
cluster radial cut--off.

We expect some radial dependence of the blue fraction because of the well known colour
segregation in clusters. At lower redshift, Fairley et al. (2002) and
Kodama \& Bower (2001) show that the blue fraction radial profiles differ from cluster
to cluster. BO show that the $f_b$ radial profile depends on redshift.
It is therefore dangerous to assume a universal $f_b$ radial profile for
all clusters. At the higher redshifts of our sample, by taking $2 \times R_{30}$
instead of $R_{30}$, the blue fraction of two clusters stays constant within 1 $\sigma$
(Table \ref{tab:bo}), whereas the blue fraction increases for cl2245-3954. 

Margoniner \& de Carvalho (2000) and Margoniner et al. (2001) (MC \& M,
hereafter) calculate
richness and blue fractions inside a fixed metric diameter of 0.7 Mpc
(for $H_0=67$ km s$^{-1}$ Mpc$^{-1}$). 
This choice is fundamentally different from the one adopted by
BO, whose metric radius scales with the cluster size. 
Because of the existence of the
morphological segregation (and hence of a colour segregation), the blue
fractions of clusters of different sizes are not directly
comparable. 

In order to remove the obvious $f_b$ dependence on cluster size, 
MC \& M parametrize it with a power law as a function of the cluster
richness, i.e. they assume that scale, richness and colour segregation
are tightly correlated, and that the correlation is the same for all
clusters and does not depend on $z$. 
In the light of the results quoted above, such an assumption
seems debatable.

As a test case, let us consider our two clusters 
cl0048-2942 and cl2245-3954. Measuring their $f_b$ and richness according to MC \& C
prescriptions, we obtain for both clusters the same richness 
($N_{MC \& C}\sim30$), whereas our asymptotic richness of the two clusters
differs by a factor of two (see Table 2). In order to obtain
the $f_b$ values derived according to the BO prescription, we need to correct the
$f_b$ derived in a 0.7 Mpc diameter
by -0.16 and +0.24, respectively, whereas the correction, according
to MC \& C prescription should be the same. 

The remaining cluster, cl2249-3958,  is contaminated in its outskirts by the presence
of a foreground group, and therefore its MC \& C richness is overestimated: the
statistical correction  adopted by them removes only the average background, whereas the
radial profile suggested by BO helps a lot in detecting groups superposed on the
cluster line of sight. This is not a rare situation, and if the measured richness is
incorrect, the richness--dependent correction is incorrect too.

The Margoniner et al. (2001) claim that richer clusters
tend to have lower $f_b$ is therefore a simple
restatement of the morphology--radius relation: the larger and richer
the cluster is, the lower is its spiral fraction (and therefore the
blue fraction) in a fixed metric aperture. This is not informative at
all on the dependence between richness and $f_b$ (as defined
by BO and/or Abell).

A similar fixed aperture for measuring the blue fraction has been
recently used by de Propris et al. (2003) and by Goto et al. (2003), 
so similar concern applies to their work. 

\subsection{Colour cuts}

A survey of the colour cuts adopted in the literature  shows different choices. Kodama
\& Bower (2001) adopted colour cut is $\Delta(g-r)=0.26$ to $0.40$ mag, depending on
redshift, whereas Ellingson et al. (2001) adopted $\Delta(g-r)=0.21$ to $0.28$ mag
instead for the very same data (filters and clusters). Inspection of Fukugita
et al. (1995) tables, used by Kodama \& Bower (2001) in their calculations, 
seems to confirm Ellingson et al. (2001) results. 

Margoniner et al. claim that $\Delta(g-r)=0.2$ mag is
equivalent to $\Delta(B-V)=0.2$ mag, and adopt the former cut for
their work. However, at $z=0$, $\Delta(B-V)=0.2$ mag is the colour
difference between (spectrophotometric) E and Sab galaxy types
(Fukugita, Shimasaku, \& Ichikawa 1995), whereas $\Delta(g-r)=0.2$ mag
is the difference in colour between E and Scd. 
By choosing $\Delta(g-r)=0.2$ mag, there are
spectrophotometric types that will be counted as red by Margoniner et
al. but as blue by BO.  Since the evolution of galaxies in clusters
seems to be rather different between early- and late-type spirals
(Dressler et al. 1997), MC \& M are sampling a population of
galaxies different from the one sampled by BO.

For our three higher redshift clusters,
by adopting $\Delta(g-r)=0.2$ mag (instead of $\Delta(B-V)=0.2$ mag
as BO prescribe) lowers blue fractions by about
0.1, simply because spirals of intermediate spectrophotometric types
are now counted as red. It is therefore clear that blue fractions
computed assuming different colour cuts cannot be directly compared.

The different MC \& M colour cut choice
and the different aperture adopted produce, as a final effect, a
$\Delta f_b$ between the two methods that can be as large as 0.4,
for an average $f_b$ of 0.2.

In a series of papers on the BO effect, Rakos and collaborators
(e.g. Rakos \& Schombert 1995; Rakos, Schombert, \& Kreidl 1991;
Steindling, Brosch, \& Rakos 2001) adopt a different definition for
the reference colour: their colour offset is measured relatively to the
colour of a present day elliptical, instead of using as reference the
observed colour--magnitude relation. In this latter case, the colour is
observed to evolve with redshift from the (redder) location of
present--day ellipticals (e.g.  Stanford, Eisenhardt, \& Dickinson
1998); this is expected since the age of the Universe at a given
redshift is an upper limit to the age of the stars at that
redshift. In other terms, their offset is not given with respect to
the colour of passively evolving objects having the age of the Universe
at that redshift (as in the BO prescription), but with respect to 15
Gyr old galaxies (even in a Universe that may be only 7 Gyr old, for
example). With Rakos et al.'s choice, a cluster composed exclusively
of passively evolving galaxies naturally increases its blue fraction
with look--back time (i.e. $z$). At high enough redshift, their colour
cut will eventually include in the blue fraction even the reddest
galaxies at that redshift. In fact, this ``high enough'' redshift
corresponds to the redshift of our clusters, the
reddest galaxies of which are almost 0.2 mag bluer than the 
colour of present--day ellipticals, thus qualifying to be classified ``blue" 
by the Rakos et
al. criterion. While their choice is fully auto-consistent, their $f_b$
cannot be directly compared with the BO one, because the two
definitions are equal only at $z\sim0$.  In particular, one needs to
de--emphasize the Rakos et al. claim that at $z\sim0.9$ almost all
galaxies are blue: this is a consequence of their definition of
``blue" since at $z=0.9$ no galaxy can be red enough to be classified
as red by their criterion, simply due to the reduced age of the
Universe at such an epoch. For this reason, it is preferable to adopt
the BO definition of the blue fraction, as we did in our analysis,
which separates the bluing
due to the young age of the Universe from the bluing due to the BO
effect itself.

Given these considerations, we believe that Rakos et al.'s large blue
fraction at high redshift is no longer in contradiction with the low
blue fraction at high redshift found by van Dokkum et al. (2000), and, 
in a more general way, their claim of a clear evidence of a BO effect 
should be considered with caution.

\subsection{Cluster selection bias}

Thus far, we have assumed that any observed sample (ours and those of
other authors) is a representative sample, i.e. that selection
criteria, if present, are benign.  Andreon \& Ettori (1999) have shown
that, instead, for the BO sample this is not the case, and that the
high redshift clusters they studied are not the ancestors of their
present-day clusters. In other words, and for that particular sample,
one is comparing ``unripe apples to ripe oranges'' in order to
understand ``how fruit ripens'' (Andreon \& Ettori 1999). The three
high redshift clusters analysed in this paper are much poorer than the
high redshift clusters in the BO sample, thus being more similar, from
this point of view, to BO's low redshift clusters. On the other hand,
they are much smaller (their $R_{30}$ is smaller) than the large
majority of the
clusters in the BO sample, and so they possibly consist in another
class of clusters with respect to the ones gathered in the BO sample.

The three clusters we study in the present paper have been
optically selected. As explained in Aragon-Salamanca et al. (1993) 
and in Andreon \& Ettori (1999), among
all clusters of a given mass, an optical selection favours those with
an unusual population of star forming galaxies. Since selection
criteria and band--shift effects become more and more important as
redshift increases, the optical selection might artificially increase
the blue fraction as the redshift increases, hence mimicking the BO
effect. With our data, we can test whether our cluster selection is
biased by such an unusual population. 

From the models
(Bruzual \& Charlot 1993) we expect that most of the cluster blue
galaxies, unless they are forming stars at a significant rate,
fall below the limiting magnitude of $m_I=22$ used by Lobo
et al. (2000) in the process of detecting clusters, ie. we would
expect them to be fainter than that limiting magnitude. 
In order to mimick a selection less biased by galaxies with
an high star formation rate, we remove the blue
galaxies from our three clusters, and we measure the detectability
of the red population by re-inserted them
in the EIS catalogue at various positions. 
As we do not know the membership of each individual red galaxy in the cluster
line of sight, we did 101 realizations of each 
background--subtracted cluster, each one using a background 
sample that is randomly extracted from our control field sample.
Each cluster was then inserted 10 times (for a total of 1010 simulations per
cluster) in the EIS catalogue, avoiding areas where clusters are detected. 
We recovered the inserted clusters 33, 50 and 65\% of
the times for cl2245-3954, cl2249-3958 and cl0048-2942,
respectively. For the detection of these three clusters, and
especially of the former two, the presence of an important (over the
whole cluster) blue population turns out to be essential, because 
clusters similar to those inserted (but without a blue population)
are underrepresented by a factor of 2 (=3/(0.33+0.50+0.65)). 
Therefore, the simple fact of having selected the sample from optical
photometry seems to have biased the blue fraction (toward higher
values observable at higher-$z$) by preferentially selecting, among
all possible clusters, the ones with a larger blue fraction, that are
more easily detected.

We remind the reader that our clusters are quite small with respect to the ones
listed in BO (see Figure 6). Their detectability would be even lower,
if they were of median size, for the
same richness. The small $R_{30}$ radii of our clusters, given their
richness, is a further reason to believe that the detected clusters
are just the tip of the iceberg.

The same concerns may apply to other clusters selected in similar
ways.

\subsection{$f_b$ errors}

Last, but not least, to conclude our discussion on the effects
that have a bearing on the detection of the BO effect, we should not
neglect the way errors are defined.

The error associated to the computed fraction of blue galaxies
strongly depends on the error definition: should or should it not take
into account the fact that we observe $n$ cluster member galaxies
subject to Poissonian fluctuations? If the aim is to measure the error
on the blue fraction of the parent distribution from which the
observed cluster is a member, then the answer is yes, it should. On
the other hand, if one restricts oneself to measuring the error on the
blue fraction of one specific cluster, then this is not needed. This fact
becomes clear by reading Gehrels (1986) that we quote here: ``We
consider [...] the case where an observer is measuring two different
kinds of distinguishable events. It is assumed that [...] the number
of events of each type [...] is distributed according to Poissonian
statistics. The objective is to obtain confidence limits on the ratio
of the two event rates based on the measurement of a small number of
events", i.e. errors for the $f_b$ fraction in our case. As Gehrels
(1986) shows, the joint probability to observe $n_1$ red galaxies and
$n_2$ blue galaxies is equal to the Poisson probability of observing
$n_1+n_2$ galaxies times the binomial probability for obtaining
specifically $n_1$ red and $n_2$ blue galaxies, given that the
combined number of galaxies observed is $n_1+n_2$. The binomial
probability alone, adopted eg. by Metevier, Romer \& Ulmer
(2000)\footnote{There is a typo in the formula quoting the variance of
binomial distribution given in Metevier, Romer \& Ulmer (2000) and in
``Notes on statistics for physicists, revised" by J. Orear, the latter
distributed by NED. The right formula for the square root of the
variance of a binomial distribution is: $\sigma(p)=\sqrt{p*(1-p)/N}$,
where $N$ is the number of trials, and $p$ is the probability of
success.}, gives errors that are somewhat conservative for small
$n_1+n_2$.

Such a discussion would be useless if the error coming from other
sources (the background subtraction, for example) dominated the
overall error budget. This is not, indeed, the case for the BO
clusters, for which the full error (including
everything) listed in BO is often smaller than the Gehrels error
alone. 


If one adopts the Gehrels' (1986) error formula,
then the error on the blue fraction of cl2245-3954
becomes non--zero, as it should be. 

To summarize, BO and - given the absence of similar remarks
in the literature - most of the literature seem to quote errors by
assuming that ``repetition of experiment'' means ``re-observing the
same cluster'' ie. observing exactly the same number of
galaxies. Here, instead, we claim that errors should be quoted as if
``repetition of experiment'' means ``observation of any cluster drawn
from the same parent population''.

The underestimate of the error implies that any existing trend of the
blue fraction with $z$ has been emphasized more than the statistics
allow. Figure 8 shows the blue fraction as a function of redshift,
plotted with error bars that now take into account the binomial term,
the Poissonian term, and over--Poissonian background subtraction
errors. For the BO clusters, we simply add (quadratically) to the
error quoted by BO the ones computed according to Gehrels (1986)
because of lack of information. The Figure also
includes our three clusters, shown as solid dots. Given the
large error bars, the data can, after all, be described by a constant $f_b$.

\begin{figure}
\centerline{%
\psfig{figure=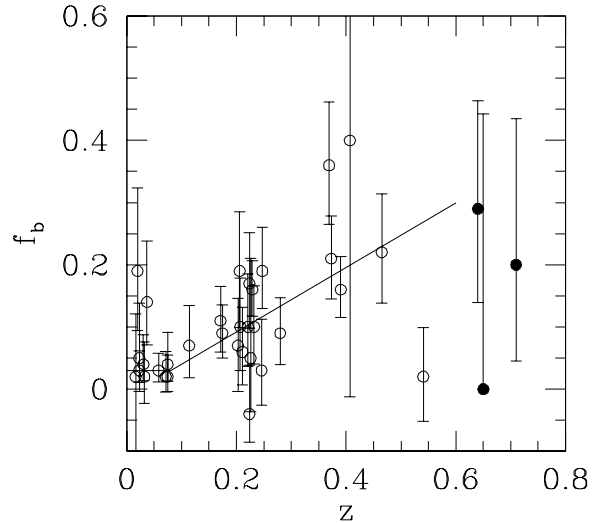,width=8truecm}}
\caption[h]{Blue fraction as a function of redshift, including errors 
on the sample representativity. Line, open and filled circles are defined as
in Figure 7.}
\end{figure}

\section{Summary}

Through the observations of three clusters at $z\sim0.7$, of richness
comparable to the low redshift sample of Butcher \& Oemler (1984; BO),
we have determined their fraction of blue galaxies, $f_b$. 
According to the standard analyses (those presented in BO,
of widespread acceptance and shown in Figure 7),
two clusters have a
low blue fraction for their redshifts, and the fraction of the
third one is compatible with the
expected value.

We studied the impact of relaxing each one of the BO criteria in the
calculation of the blue fraction.

-- The exact choice of the luminosity cut is not critical, provided it
differs by 1 mag or less. 

-- Adopting a unique metric radii for all clusters, regardless of them
being large or small, and eventually correcting for the richness
dependence as is sometimes performed in the literature, is not
informative on the BO effect because of the contamination by the colour
segregation in clusters. The
$f_b$ measured within a metric aperture is therefore informative about
something different from the BO effect.

-- The colour cut is also important. In some cases, we are unable to
reproduce colour cuts of other authors; in other cases we show that the
adopted colour cuts differ from those defined in BO. Since galaxies of
different colours have probably different star formation histories,
then the comparison of blue fractions derived using different cuts is
not straightforward.  For our three clusters, the blue fraction
decreases by 0.1 when adopting the more liberal $\Delta(g-r)=0.2$
colour cut, instead of the standard $\Delta(B-V)=0.2$.

-- The adoption of a correct reference colour is a critical point.
If the reference colour does not change appropriately with
look--back time, then the measurements of the BO effect are
contaminated by the bluing due to the younger galaxy ages at higher
redshift. Two of our clusters at
$z\sim0.7$ have a huge blue fraction according to Rakos et al.'s
definition, simply because the Universe is so young at $z\sim0.7$ that
no stellar population can be red enough to be called red according to
their criterion. However, this has nothing to do with the BO effect.

Finally,  a re--analysis of the error computations usually performed in the
literature shows that the $f_b$ errors quoted by BO (and likely by other
authors) underestimate the real errors. If we plot the original BO
data together with our three high redshift clusters, with the newly
determined error bars, we cannot exclude a constant $f_b$.

Therefore we conclude that: 

-- the correct comparison of BO effect determinations reported by different
authors is a task as difficult as performing the measurement itself and
both should be done with extreme carefulness.

-- Twenty years after the original intuition by Butcher \& Oemler,
we are still in the process of ascertaining the reality of the BO
effect.

In a future paper we will present a BO--style analysis for
an X--ray selected sample of clusters collected by the XMM--LSS project
(Pierre et al. 2003), hence overcoming the biases of the cluster
optical selection.

\section*{Acknowledgments}
 
We warmly thank Reinaldo de Carvalho for numerous discussions on the
subject. We acknowledge partial support from project
ESO/PRO/15130/1999 (FCT, Portugal).


\end{document}